\documentclass[%
reprint,
 amsmath,amssymb,
 aps,
showkeys 
]{revtex4-2}

\usepackage{packages}

\begin{document}

\title{Nonlinear response of Silicon Photonics microresonators for reservoir computing neural network}

\author{Emiliano Staffoli}
\author{Davide Bazzanella}%
\author{Stefano Biasi}%
\author{Giovanni Donati}%
\author{Mattia Mancinelli}%
\author{Paolo Bettotti}%
\author{Lorenzo Pavesi}%
\affiliation{%
Nanoscience Laboratory, Department of Physics, University of Trento, Italy.
}%

\date{\today}

\begin{abstract}
Nowadays, Information Photonics is extensively studied and sees applications in many fields. The interest in this breakthrough technology is mainly stimulated by the possibility of achieving real-time data processing for high-bandwidth applications, still implemented through small-footprint devices that would allow for breaking the limit imposed by Moore's law. One potential breakthrough implementation of information photonics is via integrated photonic circuits. Within this approach, the most suitable computational scheme is achieved by integrated photonic neural networks. In this chapter, we provide a review of one possible way to implement a neural network by using silicon photonics. Specifically, we review the work we performed at the Nanoscience Laboratory of the University of Trento. We present methodologies, results, and future challenges about a delayed complex perceptron for fast data processing, a microring resonator exploiting nonlinear dynamics for a reservoir computing approach, and a microring resonator with the addition of a feedback delay loop for time series processing.

\end{abstract}

\maketitle

\section{Introduction}

The interest in Artificial Neural Networks (ANNs) has considerably increased in recent years due to their versatility, which allows for dealing with a huge class of problems \cite{Genty2020ultrafast}. Nowadays, ANNs are mostly implemented on electronic circuits, in particular, on von Neumann architectures in their different specifications such as the general purposes CPU (Central Processing Units), the massively parallel GPU (Graphical Processing Units) or the specialized integrated circuits used to accelerate specific task such as the TPU (Tensor Processing Units) \cite{sze2017efficient,bhattacharya2021dnns,dhilleswararao2022efficient}. Very-large-scale ANN models have been elaborated which outperform human minds in given tasks \cite{meta2022human,li2022competition} at the expense of large training times and huge power consumption \cite{strubell2019energy,wu2022sustainable,boahen2022dendrocentric}. Other intrinsic limits of electronic ANNs are related, for example, to the ease in interference between electrical signals, the difficulty in handling a large number of floating point operations and a low parallel computing efficiency \cite{Sui2020review,Liu2021research,Porte2021complete}. \\

\begin{figure}[h!]
	\centering
	\includegraphics[width=0.5\textwidth]{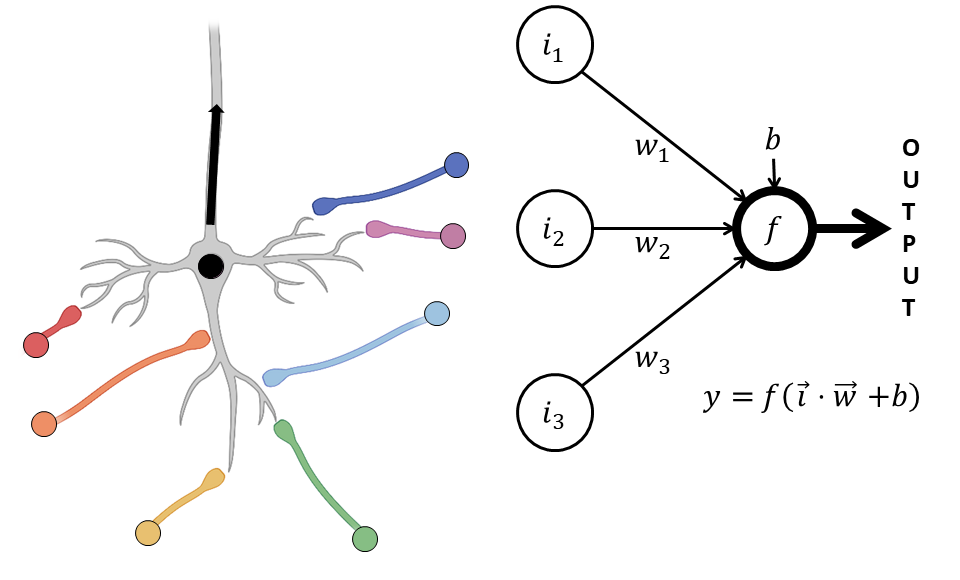}
	\caption{Left: Sketch of a biological neuron (black dot). Different signals from nearby neurons (colored) are collected by the neuronal dendrites through interconnecting synapses. The neuronal body integrates the signals and, if above a threshold, produces a voltage spike which is sent via the neuronal axons (black arrow) to the post-synaptic neurons. Right: Sketch of an artificial neuron where the output y is produced by the formula given in the inset from the different inputs $i$ (image courtesy of Gianmarco Zanardi).}
	\label{fig:opticalNeuron}
\end{figure}

A possible solution to these limitations is provided by Photonic Neural Networks (PNNs) which enable high-speed, parallel transmission (Wavelength Division Multiplexing, WDM) and low power dissipation \cite{Liu2021research,Sui2020review}. PNNs have the same overall architecture as an ANN, namely, they are made by several interconnected neurons where each neuron receives multiple inputs and feeds multiple other neurons (Fig. \ref{fig:opticalNeuron}). The received inputs are weighted, combined, and processed by each neuron which through a nonlinear activation function feeds its interconnected neurons.  When optics comes into play, some of these operations are very easy to implement. For example, large matrix multiplication becomes very fast and energy-efficient \cite{cheng2021photonic}, giving PNNs a great advantage compared to electronic ANNs. These advantages lead to the development of photonic accelerators for electronic ANNs \cite{zhou2022photonic}. On the other hand, in integrated PNNs, the inter- and intra-neurons connections are easily established through waveguides in an on-chip optical switching network \cite{Sui2020review,testa2019integrated} where the propagating optical signal can be modified using tunable waveguide elements (e.g. phase shifters or Mach Zehnder interferometers MZI \cite{vivien2016handbook}).\\

The complexity in operations achievable by a PNN depends on multiple factors. These include the network topology, namely how the single processing units (neurons or nodes) are interconnected. Artificial neurons are combined in a huge variety of networks, from basic structures (e.g. the single perceptron shown in Fig. \ref{fig:opticalNeuron} right \cite{Rosenblatt1958perceptron}) to very complex ones \cite{shakrawy2020Principle}. Potentially, any topological organization of neurons can be achieved, the optimal structure depending on the specific task to be solved and on the amount and format of data to be analyzed \cite{brunner2021competitive}. Problems that require low latency and fast reconfigurability fit with PNNs of the feed-forward type, where the data flows only in one direction (from the input neurons to the output neurons) through different layers of nodes \cite{pai2020parallel}. On the contrary, high-complexity tasks where long short-term memory plays a key role require recurrence where the information flow back and forth between the neurons \cite{medsker1999recurrent}. A model which is easily implemented in PNNs is the photonic reservoir computing, where random fixed connections between the nodes are established, with training only performed in the output layer \cite{antokik2020large}. Finally, the readout strategy is another key element since it provides direct access to the information elaborated by the network itself. The readout can be optical or electrical and its choice again depends on the topology of the network and the specific requirements of the task \cite{ma2021comparing}.\\

In ANNs, the nonlinear activation function is implemented within the node. This aspect of the neuron plays a fundamental role in the learning process since it determines the output of the node. Here, PNNs provide many possible choices \cite{dabos2022neuromorphic,el2022photonic}. For example, a suitable activation function is the square modulus $\vert \cdot \vert^2$: this is easily implemented by a direct detection process of the optical signal (e.g. with a photodiode) \cite{hamerly2019large,shen2017deep}. The intensity (and thus the power) $I(t)$ associated with an optical signal is indeed directly proportional to the square modulus of the electric field $E(t)$, i.e. $I(t) \propto \vert E(t) \vert^2$ \cite{agrawal2002receivers}. 
Another activation function can be provided by a Semiconductor Optical Amplifier (SOA) integrated within the neuron \cite{shi2019deep}. An SOA behaves linearly for low input optical power but reaches saturation for higher power values \cite{agrawal2002amplifiers}. Its power-gain curve thus is strongly nonlinear, making SOAs suitable for acting as a nonlinear node in a PNN. Here, we are mostly interested to discuss nonlinear nodes based on microring resonators \cite{bogaerts2012silicon,pavesi2021thirty}. Since the microring's optical transfer function depends on the stored optical power \cite{Borghi2021modeling}, they can be used to implement different kinds of nonlinear transfer functions \cite{Biasi2023}.\\

The optical domain offers an optimal testing ground for ANNs that process information using complex-valued parameters and variables \cite{lee2022complex}. The light propagation in waveguides and its nonlinear interaction with various media are naturally described in the complex domain where both the phase and the amplitude of the electric field associated with the optical signal have to be taken into consideration. Complex numbers are thus intrinsically involved in optical systems which turns out to be ideal for the implementation of complex-valued neural networks \cite{hirose1994application,lee2022complex,hirose2012generalization}. Even though each complex number can be represented by means of two real numbers, a complex-valued ANN must not be considered equivalent to a real ANN with a doubled number of parameters \cite{hirose1994application,hirose2012generalization}. Indeed, when it comes to complex multiplication, the rotatory dynamics of complex numbers enter into play, leading to a reduction of the degrees of freedom as compared to the case of completely independent parameters. Therefore, this opens further opportunities for PNNs which easily manipulate complex numbers. This possibility, associated with properly chosen nonlinear nodes and an effective readout strategy, yields that even a simple hardware implementation of a PNN manages to perform demanding tasks, which would require a much higher cost if faced with traditional ANNs \cite{mancinelli2022perceptron}.

In this chapter, we will discuss a few simple PNNs implemented on a silicon photonics platform \cite{pavesi2021thirty}, that demonstrate the basic mechanism of silicon-based PNNs. Silicon photonics is particularly interesting since its easy integration with electronics allows for on-chip training of the network and for volume fabrication of the PNNs \cite{vivien2016handbook}. In section \ref{sec:II}, a simple optical neuron is discussed where different delayed versions of the input optical signal are made to interfere before the output port \cite{mancinelli2022perceptron,staffoli2023}. In section \ref{sec:III}, the simple microring resonator is used to demonstrate complex nonlinear dynamics \cite{Borghi2021modeling}. In section \ref{sec:IV}, a Reservoir Computing network implemented by a single microring resonator within a time delay scheme is used for complex classification tasks \cite{borghi2021reservoir}. In section \ref{sec:V}, linear and nonlinear memory tasks are used to evaluate the memory capacity of a microring resonator \cite{bazzanella2022microring}. Section \ref{sec:VI} shows the possibility to extend the microring resonator fading memory by using an external optical feedback loop \cite{Donati22microring}. Finally, section \ref{sec:VII} concludes the chapter with a summary and perspectives.

\section{A simple photonic network: the delayed complex perceptron}
\label{sec:II}

The simplest perceptron consists of an algorithm that associates to two given input and weights vectors, respectively $\Vec{x}$ and $\Vec{w}$, the output of an activation function $f(\Vec{x} \cdot \Vec{w} )$, according to \cite{Rosenblatt1958perceptron}

\begin{equation}
    f(\Vec{x} \cdot \Vec{w}) =    
    \begin{cases}
        1   \qquad \text{if} \quad \Vec{x} \cdot \Vec{w} > 0 ; \\
        0   \qquad \text{otherwise} ;
    \end{cases}
\end{equation}

where $\cdot$ is the inner product of the Euclidean space. It can be considered a binary classifier without memory. Still, it can be used to describe the working principle of the individual nodes in complex topologies. 

\begin{figure}[h!]
    \centering
    \includegraphics[width=0.5\textwidth]{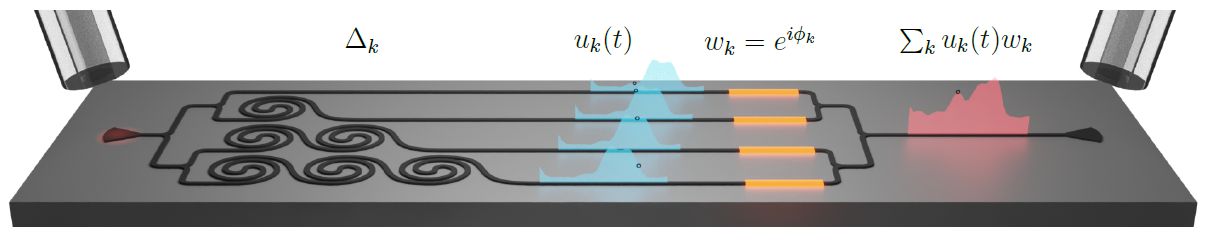}
    \caption{Sketch of the integrated photonic circuit which performs as a complex perceptron. Gratings are used to couple the light in and out of the optical circuit. The red spot on the left shows the input signal. A $1\times 4$ splitter distributes the input signal to four delay lines, realized by spirals (each spiral adds a delay of $\Delta_t=50$ ps). Then, thermal phase modulators (yellow segments) allow controlling the relative phases ($w_k = e^{i\phi_k}$) of the signals ($u_k(t)$, blue lineshapes). These are then summed by combiners and the resulting signal (red lineshape) is output via a grating. A fast detector provides the output nonlinear node and yields the output signal $y(t)$. Adapted from \cite{mancinelli2022perceptron}.}
    \label{fig:cperc_scheme}
\end{figure}

A modified version of this simple algorithm has been implemented in an optical circuit to realize what we named a Delayed Complex Perceptron (DCP) \cite{mancinelli2022perceptron}. Its structure is illustrated in Fig. \ref{fig:cperc_scheme}. The input optical signal ($u(t)$) is split into 4 channels ($u_k(t), k=1, \dots, 4$), where the waveguides are spiralized so that a delay multiple of $\Delta_t = 50$ ps is induced with respect to an unperturbed copy traveling in the top channel ($u_1(t)$). The phase of the signal in each channel is then modified through independent phase shifters, which are actuated by micro-heaters (in yellow in Fig. \ref{fig:cperc_scheme}). Therefore, the relative phase $\phi_k$ of each signal can be controlled by the driving current in the phase-shifters and acts as weight $w_k=e^{i\phi_k}$. Indeed, these currents represent the tunable parameters in the network during the learning phase. Finally, the modified signals are made to interfere (summed) in the output combiner. The result of this interference provides the output optical signal. The nonlinear node of the PNN is here represented by a fast photodiode that detects the output signal intensity $y(t)$ by performing the square of the output signal. The ultimate purpose of the DCP is to combine the information of the input signal at the present time and at fixed delays in the past. The role of the phases is to modulate the interference between the signals in the different channels, selecting thus the proper combination of information from each time instant to perform the assigned task. The algorithm which describes the DCP is

\begin{equation}
	y(t)= f(\Vec{x}\cdot\Vec{w})=\left|\sum_{k}u_k(t)e^{i\phi_k}\right|^2.
\end{equation}

The DCP is fabricated on a Silicon-on-insulator (SOI) wafer, being the silicon layer 220 nm thick. The waveguides are 450 nm wide which allows for single-mode operation on the TE (transverse electric) polarization fixed by the input grating coupler. \\

The classifier nature of the perceptron can be declined into different tasks. For example, the DCP has proven effective in solving logical tasks involving two bits separated in time (e.g. the XOR performed between the current bit and the first in the past) \cite{mancinelli2022perceptron}. In another application, the DCP is used as a compensator for distortions induced by the chromatic dispersion on optical signals propagating in fiber \cite{staffoli2023}. In fact, chromatic dispersion causes an intersymbol interference between adjacent bits, with the consequent loss of information at the receiver \cite{agrawal2002fibers}. Nowadays, the recovery process can be accomplished by Dispersion Compensating Fibers (DCFs) \cite{Vengsarkar1993dispersion}, which however are non-tunable devices and introduce latency \cite{Granot2020fundamental}. An alternative is represented by Bragg Gratings \cite{Ouellette1987dispersion}, but, since they work on the entire WDM aggregate, they cannot perfectly compensate all the channels. The DCP constitutes an alternative to these methods, with the advantage of being reconfigurable and providing a drastic latency reduction. Moreover, compared to other technologies that implement coherent receivers and digital signal processing (DSP) for equalization, the DCP has the advantage of operating the corrections directly on the optical sequence, avoiding the complexity and the energy demand of DSP (which represent a significant fraction of the power budget).\\

\begin{figure}[h!]
    \centering
    \includegraphics[width=0.5\textwidth]{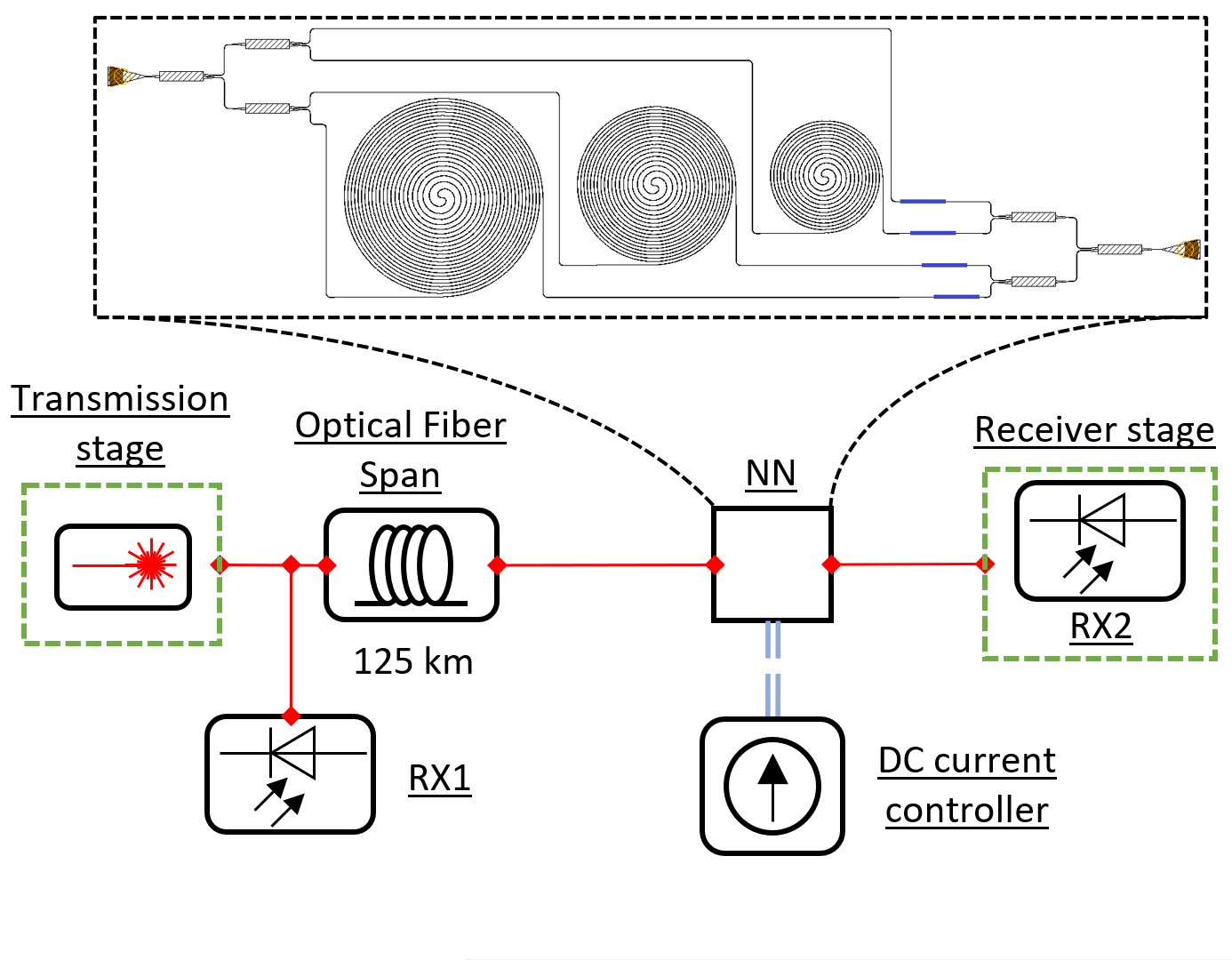}
    \caption{Experimental setup. Light is generated and modulated in the transmission stage and then sent into a 125 km optical fiber span. It proceeds then through the DCP for optical processing. Two fast photodiodes monitor the input (RX1) and the output (RX2) signals. The inset shows the actual design of the NN device, where one can observe the cascaded $1\times4$ and $4\times1$ splitter and combiner, the three spirals, and the four phase shifters (small blue rectangles) connected to the external DC current controller. Adapted from \cite{staffoli2023}.}
    \label{fig:setup}
\end{figure}

The experimental setup of Fig. \ref{fig:setup} has been used to access the compensation capabilities of the DCP. In the transmission stage, a laser source operating in the third telecom window is modulated as a 10 Gbps Non-Return-To-Zero (NRZ) signal, based on a Pseudo-Random Binary Sequence (PRBS) of order 10 and period $2^{10}$ bits. A Fiber Optic Coupler sends part of the optical power to a fast photodiode (RX1), while the other fraction proceeds into an optical fiber with a length of 125 km. The distorted signal is then coupled to the DCP for optical processing and it is finally detected by a fast photodiode (RX2). 

During the training, for each tested configuration of the injected currents, the input and output curves are acquired and compared in order to determine the expected level (0 or 1) associated with each output bit. The loss function provided to a Particle Swarm Optimizer \cite{schutte2004parallel} for the training aims to create the maximum relative separation between the distributions associated with the two classes, namely 0 and 1 (i.e. the maximum contrast between levels). This leads to a reduced Bit Error Rate (BER).

\begin{figure}[h!]
    \centering
    \includegraphics[width=0.5\textwidth]{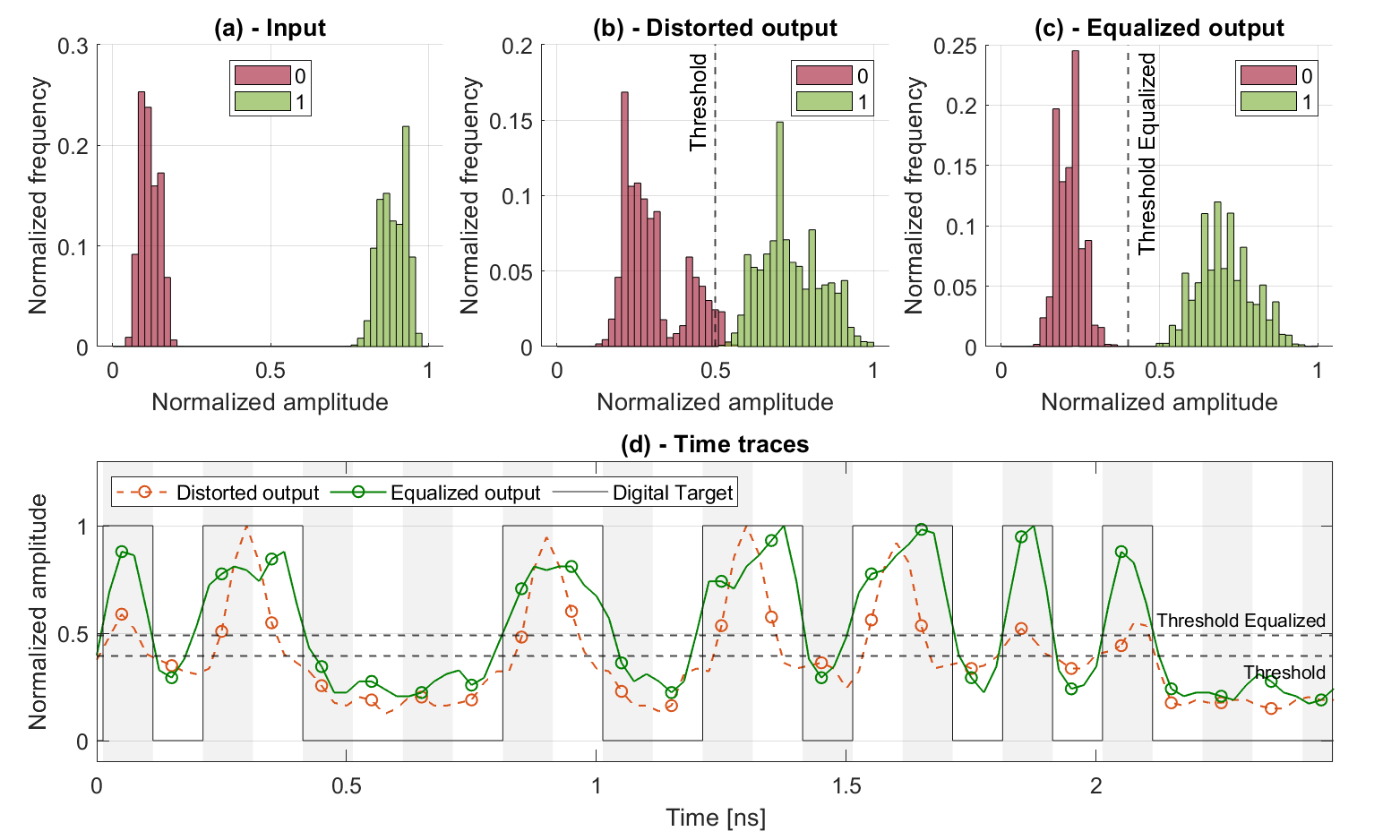}
    \caption{Results for 125 km fiber compensation. Distributions of expected 0s (red bars) and 1s (green bars) in (a) input, (b) uncompensated output, and (c) compensated output. (d) Time traces of target (black line), uncompensated (red line), and compensated output (green line). Circles represent the reference sample for each bit. Adapted from \cite{staffoli2023}.}
    \label{fig:125kmres}
\end{figure}

The compensating effect of the perceptron for a span of 125 km is summarized in Fig. \ref{fig:125kmres}. The intersymbol generated by chromatic dispersion generates the closing of the gap between the distributions in Fig. \ref{fig:125kmres}(b). The fact that the two distributions are close to each other or even overlapping leads to an increased probability of incorrect identification of the bit value when a threshold is applied, thus the BER increases. The trained DCP manages to combine information coming from the present, the first, and the second past bits to split the distributions. In this way, a BER reduction is achieved.\\

The performances of the DCP are comparable with other alternative approaches such as those in \cite{Sackesyn,Argyris2018}, which exploit photonic reservoir computing. The benefits of the DCP regard the simpler architecture, minimized latency, and full on-chip signal processing with consequent optical readout strategy (except for the training phase). Moreover, neither the response of the system is determined by random connections between nodes nor recurrence is present in the network, whose action is thus simpler to simulate. In the future, the recovery of nonlinear effects mediated by self-phase modulation is going to be attempted too, foreseeing implementations in transmission optical lines. With this perspective, in order to provide the user with a ready-to-use full-optical transceiver, the next-generation devices will be provided with a full-optical activation function stage directly in the photonic chip itself. This can be accomplished both by active components as integrated SOA with nonlinear properties and also by passive structures (e.g. integrated microring resonators) working in a nonlinear regime.

\section{Nonlinear dynamics in a microring resonator}
\label{sec:III}

Microring resonators (MRs) are resonant devices, characterized by compact footprint, and wide operation bandwidth \cite{bogaerts2012silicon}. When operating close to the resonant condition, a strong increase in power density in the MR occurs, inducing a nonlinear response. These features make them versatile photonic structures for different applications, including communications \cite{testa2019integrated}, bio-sensing \cite{steglich2019optical}, spectroscopy \cite{suh2016microresonator}, frequency metrology \cite{papp2014microresonator}, and quantum optics \cite{llewellyn2020chip}. The temporal dynamics of a MR have to consider many intertwined nonlinear processes. Sufficient light intensity triggers nonlinearities related to the Si third-order susceptibility \cite{borghi2017nonlinear}, specifically Two-Photon Absorption (TPA). TPA generates free carriers which thermalize by the emission of phonons and heat up the MR. The temperature ($T$) and the free-carrier (FC) density ($N$) are influenced by the optical power ($P$) circulating in the MR and, in turn, generate a nonlinear variation of the MR refractive index ($n(P)$):
\begin{equation}
	n(P)=n_0+\frac{dn}{dT}\Delta T-\frac{dn}{dN}\Delta N,
\label{eq:ref}
\end{equation}
where $n_0$ is the linear refractive index, $dn/dT$ the thermo-optic (TO) coefficient and $dn/dN$ the free carrier dispersion (FCD) coefficient \cite{borghi2017nonlinear}. Noteworthy, since both coefficients are positive, the TO and FCD phenomena exhibit opposite sign. The MR resonant condition ($m \lambda_{res}= 2\pi R  n(P)$, where $m$ is resonant order and $R$ the MR's radius) is thus power dependent which, in turn, alters the optical power circulating in the MR. In this way, the pattern of inter-dependencies is complete \cite{johnson2006selfinduced}. The dynamics given by the coupling between these processes are described by three coupled differential equations \cite{vaer2012simplified}, which indeed describe the variation of internal energy, FC population, and temperature. 
The complexity of the dynamics in the MR may generate temporal instability of the transmitted optical signal. Even when the system is fed with a continuous wave (CW) source, under particular conditions one may observe self-pulsing (SP), bi-stability, chaos, excitability, or a nonlinear distortion in the transmitted spectrum \cite{deCea2019power,Luo2012power,libin2014experimental,zhangfei2013multibistability,mancinelli2014chaotic,Takemura2020designs}. \\ 

\begin{figure}
    \centering
    \includegraphics[width=0.5\textwidth]{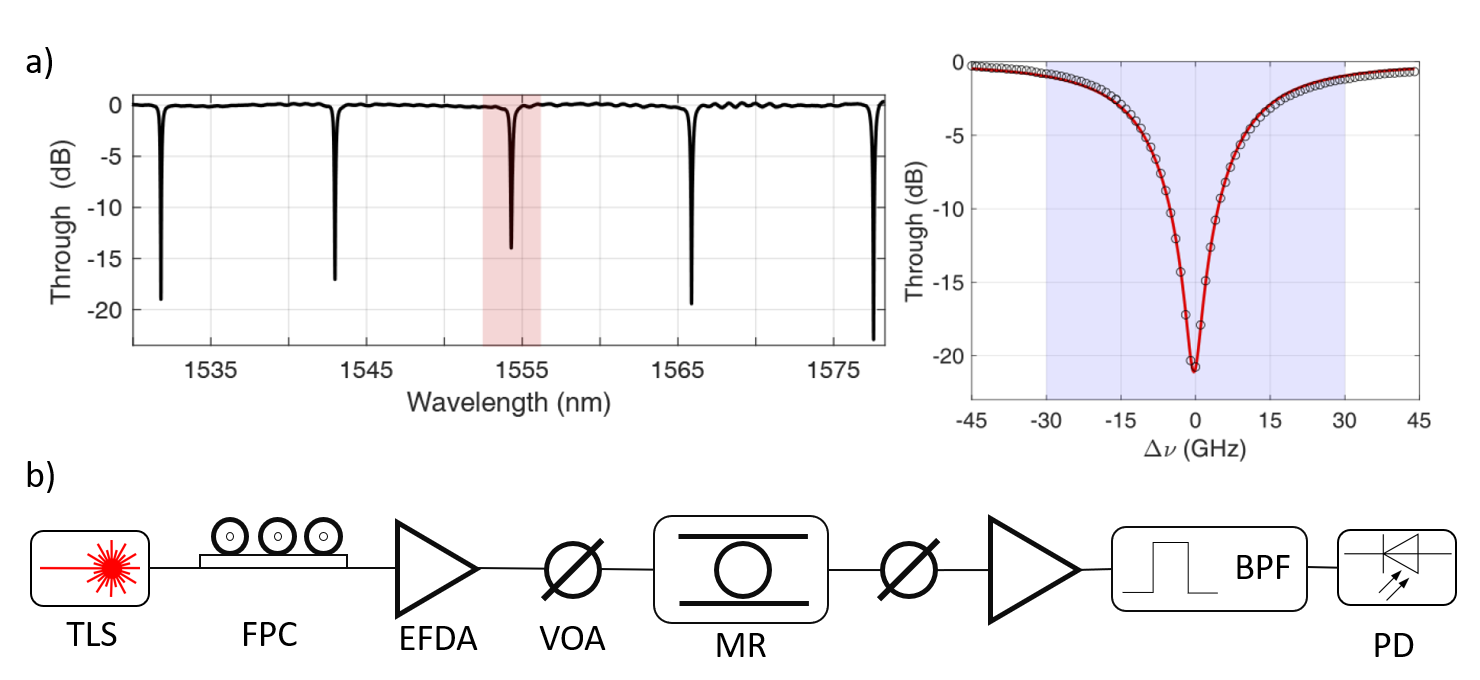}
    \caption{(a) Low power transmission spectra of the MR, collected at the Through port. Highlighted in red is the resonance order where the self-pulsing (SP) regime is analyzed. An enlarged view of this resonance is shown on the right panel. Here, the blue region marks the range of laser tuning $\Delta \nu$ used. (b) The experimental setup implemented to measure the SP of the MR and the carrier lifetime. A CW Tunable Laser Source (TLS) operating in the C-band is polarization stabilized through a Fiber Polarization Controller (FPC) and amplified by an Erbium Doped Fiber Amplifier (EDFA), before being coupled with the MR through a grating coupler. The optical power sent to the Input port of the MR is controlled through a Variable Optical Attenuator (VOA). The optical signal at the output of the Drop port of the MR is collected through another grating coupler and sent to an optical band-pass filter (BPF). Finally, the signal is detected by a 20 GHz bandwidth fast photodiode (PD) connected to a 40 GSa/s oscilloscope. Adapted from \cite{Borghi2021modeling}.}
    \label{fig:setupborghi}
\end{figure}

The insurgence of these effects in a MR fabricated on a SOI wafer has been studied in \cite{Borghi2021modeling}. A MR constituted by a 220 nm $\times$ 450 nm$^2$ Si waveguide, with a 7 $\mu$m radius in an Add-Drop configuration was used. The coupling occurs through a 250 nm gap over a length of 3 $\mu$m, corresponding to a coupling coefficient of $k^2=0.063$. The measured quality factors (Q-factors) amount to $Q_i = 1.11(8) \times 10^5$ (intrinsic) and $Q_L = 6.5(2)\times 10^3$ (loaded). The low-power transmission at the Through port is shown in Fig. \ref{fig:setupborghi}, together with the experimental setup used for the nonlinear regime of operation. 

\begin{figure}
    \centering
    \includegraphics[width=0.5\textwidth]{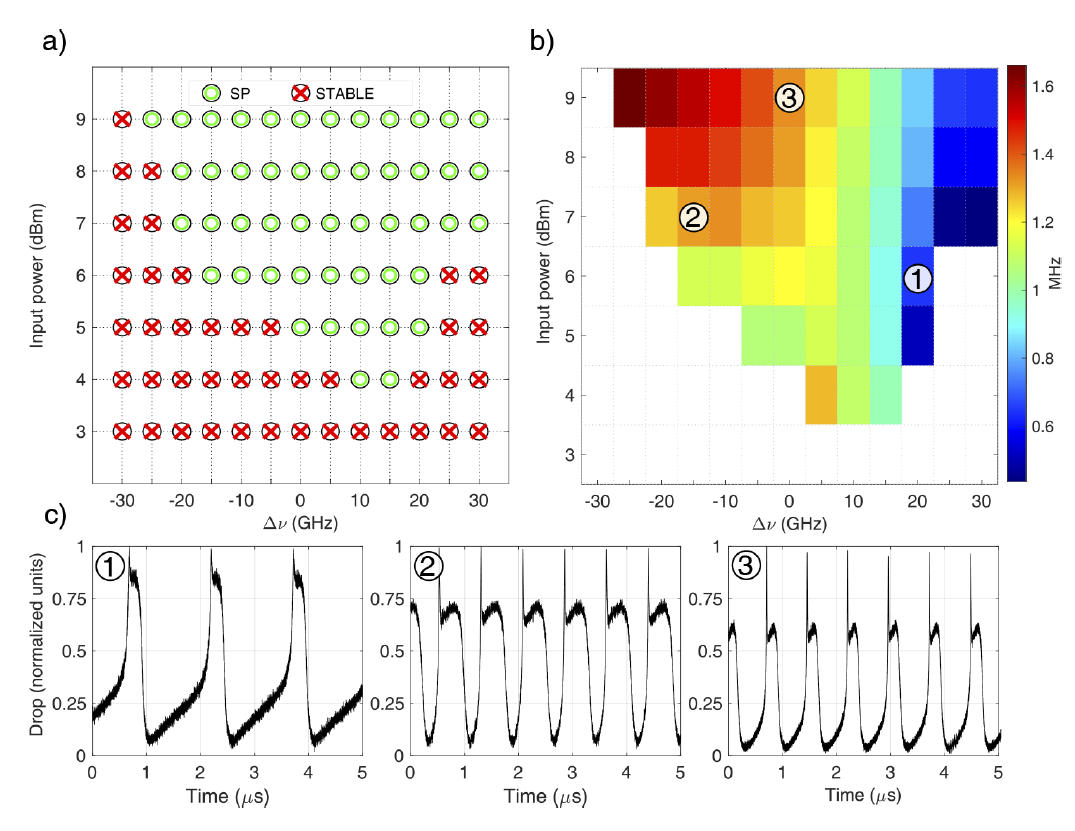}
    \caption{(a) Stability map of the MR in the $\Delta\nu$ = laser frequency detuning, $P$ = input power plane. Red crosses indicate the points where the MR, after an initial transient, shows stable output. Green circles indicate the points where the MR is self-pulsing (SP). (b) Map of the oscillation frequencies of the self-pulsing regime given in the side color code bar. (c) Examples of time traces recorded at the output of the Drop port of the MR. The maximum of the intensity is normalized to one. The labels (1), (2), and (3) refer to the values of ($P$, $\Delta\nu$) associated with each trace, which are indicated in panel (b). Figure from \cite{Borghi2021modeling}.}
    \label{fig:stabilitymap}
\end{figure}

The MR has been investigated through scans both in the input optical CW power ($P$) and laser frequency detuning ($\Delta \nu$) with respect to the cold resonance frequency of the MR itself. Figure \ref{fig:stabilitymap} presents the stability regions as a function of $P$ and $\Delta \nu$. As expected, for low input power no combination of ($P$, $\Delta\nu$) exists for which a SP phenomenon is observed, since high input power is necessary (but not sufficient) for triggering this phenomenon. Panel (b) indicates the frequency of the measured SP extrapolated from the time traces, showing the tendency to generate faster oscillations at high powers in the red-shifted detuning region. 
By looking at the time traces of panel (c), it is possible to observe how each nonlinear phenomenon contributes to the generation of SP. Each cycle starts with the generation of FC population induced by TPA, which in turn triggers free-carrier dispersion and a blue shift of the resonant frequency of the ring. All these mechanisms generate an hysteresis response, and thus bistability in the ring. The structure remains in the bistable regime (narrow peak) for a short time, due to the relaxation of free carriers, with a typical time $\tau_{fc}$. The resonance frequency is lowered, leading to a quasi-constant transmission for a short time (central region of the pulse). Finally, due to the heating of the microring generated by the free carrier relaxation, the Thermo-Optic Effect becomes predominant over the FCD, which drives the MR out of resonance (red-shift). Consequently, the Drop port transmission decreases. Then, the MR cools down with a typical time $\tau_{th}$ and gradually returns to the initial state and a new cycle begins. 

The frequency at which SP occurs depends on both $P$ and $\Delta\nu$, ranging from a minimum of $\sim 400$ kHz to a maximum of $\sim 1$ MHz. The actual frequency value roughly depends on $\tau_{fc} + \tau_{th}$, since both the relaxation dynamics have to occur to complete a full cycle of oscillation. The observation of sub-MHz self-pulsing suggests values for $\tau_{fc} \sim 45$ ns and $\tau_{th} \sim 270 $ ns, that are much larger than the typical ones for SOI waveguides \cite{libin2014experimental,pernice2010time,VanCampenhout2009silicon,Almeida2004alloptical,mancinelli2014chaotic}. Describing the internal dynamics of the MR is made even more complex by the dependence of these parameters on the instantaneous carrier concentration and the material properties \cite{Borghi2021modeling}. The commonly used approach to describe the temporal dynamics may often be reductive and the approximations behind its derivations too simplistic. The canonical three coupled differential equations may need some adjustment depending on the specific framework on which they are applied \cite{Borghi2021modeling}, but in general, they provide the physical sense behind the nonlinear processes in a MR. 
The coupling between nonlinear processes in a MR is difficult to describe, nonetheless, it is one of the key mechanisms behind the versatility of these devices. MR find wide applications in the field of PNNs too \cite{Biasi2023}, for example as memory units, optical filters, or nonlinear nodes, as will be discussed in the next sections. Their small footprint makes them suitable for integrated solutions.

\section{Silicon microring resonator for reservoir computing}
\label{sec:IV}

Particularly suitable for PNN is the Reservoir Computing model for ANN (RC-NN)\cite{antokik2020large,brunner2019photonic}. We do describe here a simple implementation of this ANN model which is based on a MR in a pump-and-probe configuration and uses the concept of virtual nodes to enlarge the complexity of the network \cite{borghi2021reservoir}.

\begin{figure}
    \centering
    \includegraphics[width=.5\textwidth]{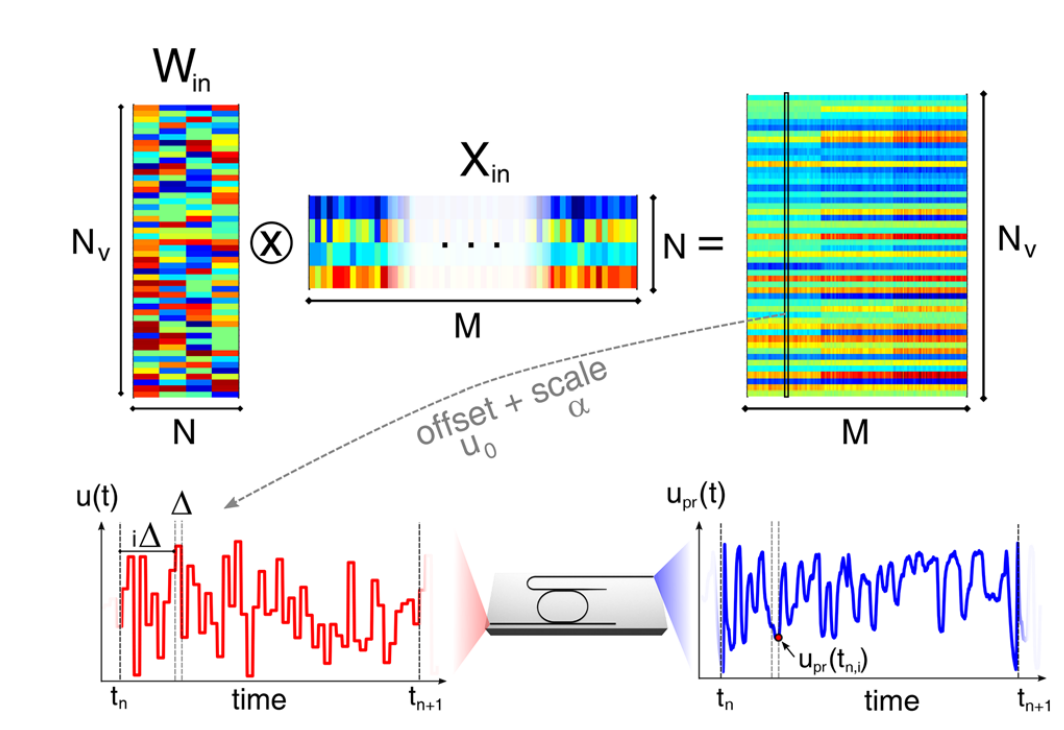}
    \caption{Process flow of the encoding of the input signal. $M$ input samples $\{ \bm{x}^{(1)},\dots,\bm{x}^{(M)} \}$ of dimension $N$ are queued on the columns of a matrix $\bm{X}_{in}$. The dimension of each sample is then increased to $N_v$ using a connectivity matrix $\bm{W}_{in}$. A global offset $u_0$ is applied to $\bm{W}_{in} \bm{X}_{in}$ to remove the negative values, and a multiplicative scale factor $\alpha$ is applied. The resulting column values represent the input pump power (red curve) $u$ of each sample, which are sequentially injected at times $t_n = n(N_v \Delta)$ at the input port of the MR (central inset). Similarly, the values of the probe power $u_{pr} (t_{n,i})$ at times $t_{n,i} = n(N_v \Delta) + i\Delta$, with $i = \{1, \dots , N_v\}$, define the virtual nodes at the output of the Drop port of the MR. Figure from \cite{borghi2021reservoir}.}
    \label{fig:modulation}
\end{figure}

The pump signal is first modulated according to the procedure described in Fig. \ref{fig:modulation}. The RC-NN receives in input a sequence  $\bm{X}_{in} = [\bm{x}^{(1)}, \dots, \bm{x}^{(M)}]$ of $M$ bits of dimension $N$. The dimension of each bit is increased to $N_v$ (which is the number of virtual nodes) by means of a proper connectivity matrix $\bm{W}_{in}$. A scale factor $\alpha$ and an offset $u_0$ are applied to the so-obtained $N_v \times M$ matrix, resulting in the sequence of operations which can be written as $ U = \alpha (\bm{W}_{in} \bm{X}_{in} + \bm{1} u_0)$. The $n$-th column $\bm{u}^{(n)}$ of the resulting $N_v \times M$ matrix codifies for the $n$-th input sample $\bm{x}^{(n)}$ and is imprinted into the pump signal $u(t)$ in the time interval $[t_n = nT, t_{n+1} = (n+1)T)$, where $T$ is the duration of an input bit. The temporal sequence is obtained by holding fixed each value in $\bm{u}^{(n)}$ for a time $\Delta$, having $T= N_v \Delta$. The average power of the modulated pump sequence $P_p$ at the input of the MR has to be kept sufficiently high to trigger nonlinearities. The pump is combined with a CW probe signal with power $P_{pr}\ll P_{p}$ and then the resulting signal is coupled with the MR, where its nonlinear dynamics transfer information from the pump to the probe. The temporal sequence $u(t)$ generates at the Drop port an output probe signal $u_{pr}(t)$. The response of the reservoir to an input sample $\bm{x}^{(n)}$ is then the sequence $\bm{u}_{pr}^{(n)} = [u_{pr}(t_n),\dots, u_{pr}(t_n + N_v \Delta)]$, where the samples represent the virtual nodes of the RC network and are acquired simultaneously with the pump signal. In the approximation of small perturbations to the input power $u(t)$ with respect to a reference value $\overline{u(t)}$, the output signal $u_{pr}(t)$ can be written as \cite{borghi2021reservoir} 
\begin{eqnarray}
\begin{split}
    u_{pr}(t) = c_0 + 
    c_1 \int_{-\infty}^{t} e^{- \left( \frac{t  -\xi}{\tau_{fc}} \right) } u^2(\xi) d\xi \\
    + c_2 \int_{-\infty}^{t} e^{- \left( \frac{t  -\xi}{\tau_{fc}} \right) } u^2(\xi) u_{pr}(\xi) d\xi \qquad,
\end{split}
\label{eq:upr}
\end{eqnarray}
where $\tau_{fc}$ is the free carrier lifetime and $c_0$, $c_1$ and $c_2$ are defined in \cite{borghi2021reservoir}. The first integral term in Eq. \ref{eq:upr} describes the intrinsic nonlinear memory of the system induced by the relaxation time of the free carriers generated with TPA. This fading memory has a duration on the order of $\sim 3 \tau_{fc}$. The second integral term describes a nonlinear coupling between the virtual nodes, which is triggered by the nonlinear free carrier dynamics in the resonator in response to variations of the pump signal. This yields recurrence and creates connectivity between the virtual nodes in the reservoir.

The RC network serves the purpose of projecting the input sequence $\bm{X}_{in}$ to a state matrix $\bm{X} = [\bm{u}_{pr}^{(1)}, \dots, \bm{u}_{pr}^{(M)}]$ in a higher dimensional space, in which the observables $\bm{Y}$ (in general an $M\times Q$ dimensional matrix) linked to the input sequence are linearly separable. In this new space, the relation between virtual nodes $\bm{X}$ and the predictions $\bm{Y}$ can be written as $\bm{Y} = \bm{W}_{out} \bm{X}$. The task then is reduced to find the output weight matrix $\bm{\Tilde{W}}_{out}$ that minimizes the regularized least square error, defined as $\sum_{k=1}^M \vert\vert (\bm{Y} - \bm{\Tilde{W}}_{out} \bm{X} \vert\vert^2 + \lambda^2 \vert\vert\bm{\Tilde{W}}_{out} \vert \vert^2$, where $\lambda$ is the regularization parameter and it is determined by a 5-fold cross-validation \cite{Tikhonov1995numerical}.\\

\begin{figure}
    \centering
    \includegraphics[width=.5\textwidth]{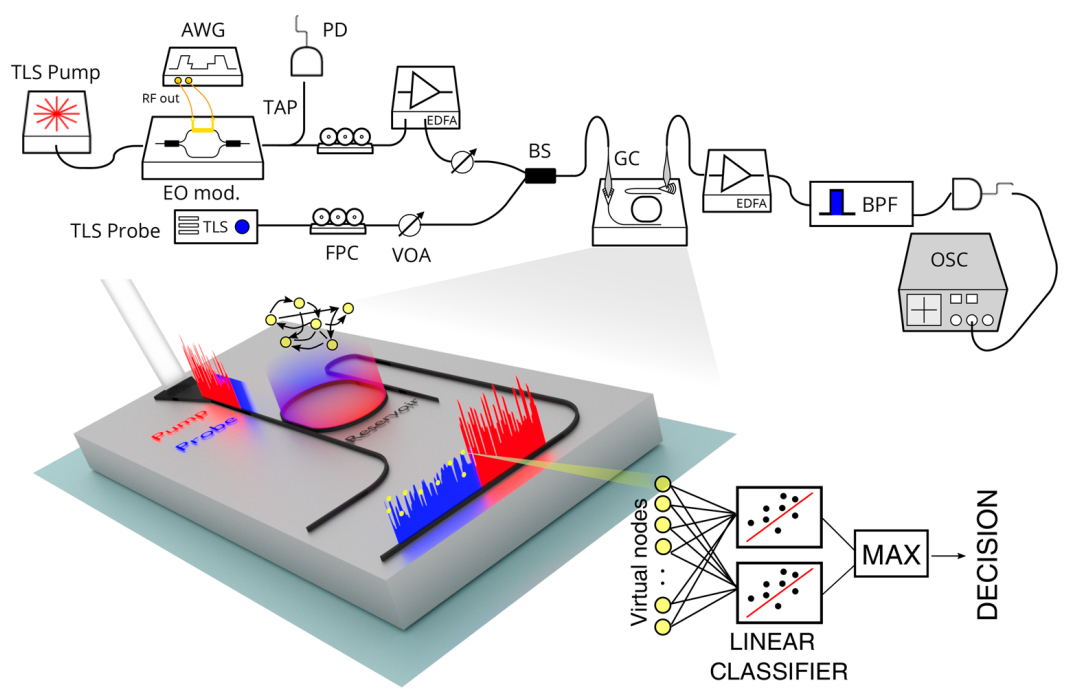}
    \caption{Sketch of the experimental setup. TLS = Tunable Laser Source, AWG = Arbitrary Waveform Generator, EO mod. = Electro Optic modulator, FPC = Fiber Polarization Controller, VOA = Variable Optical Attenuator, PD = Photodiode, EDFA = Erbium Doped Fiber Amplifier, BS = Beam Splitter, GC = Grating Coupler, BPF = Band-Pass Filter, OSC = Oscilloscope. An enlarged view of the device layout and the main logical steps which describe how the information is processed are shown in the bottom part of the figure. The intensity-modulated pump (red) and the CW probe (blue) are injected into the input GC. The incoherent transfer of information from the pump to the probe occurs within the resonator (reservoir), where the different virtual nodes (yellow dots) interact and process the input data. The probe exits from the Drop port, carrying the result of the computation. Virtual nodes are sampled and sent into several linear classifiers, each trained to recognize a specific class. The decision-making process is based on a winner takes all scheme. Figure from \cite{borghi2021reservoir}.}
    \label{fig:borghi2setup}
\end{figure}

The experimental implementation of the RC network is presented in Fig. \ref{fig:borghi2setup}. A tunable laser generates a CW pump signal, on which the desired pattern is imprinted through an Electro-Optical Modulator. The pump signal is then amplified and combined with a weak CW probe signal generated by another tunable laser. These are then coupled to the Input port of a MR through a grating coupler. The pump and probe signals are detuned by $\Delta\lambda_{p}$ and $ \Delta\lambda_{pr}$ with respect to two different cold resonant frequencies of the MR (e.g. 1549 nm and 1538 nm). Inside the MR, the magnitude of the free-carrier dynamics varies according to the modulation imprinted on the pump signal and this dynamic alters the intensity of the probe signal too due to the power dependence of the MR refractive index (Eq. \ref{eq:ref}). The output probe signal $u_{pr}(t)$ at the Drop port is recorded, sampled at the virtual nodes, and digitally processed. \\

Proof of the principle of operation of the RC network, which involves the interplay between the nonlinear transformation of the inputs and the presence of a fading memory, has been given with the 1-bit delayed XOR task. Given a binary input sequence $\bm{x}$, at time $t_k$ the RC network has to provide the result of the XOR operation applied to the input bit $x^{(k)}$ and $x^{(k-1)}$. For this particular task, the input sequence is constituted by a PRBS sequence of bit rate $B$, and the connectivity matrix $\bm{W}_{in} = [1,1,1]^{T}$ is adopted. Thus, the only mediation imposed by the mask is to bring the number of virtual nodes to $N_v = 3$, without altering the original information contained in the input sequence before this is imprinted in the pump signal. The time spacing between the virtual nodes results to be $\Delta = (B N_v)^{-1}$. The detunings are set to $\Delta \lambda_{p} = \Delta \lambda_{pr} = 60$ pm and the optical powers to $P_{p} = 3$ dBm and $P_{pr} = -3$ dBm, respectively. A threshold is applied to the outcome of the predictor $\bm{W}_{out} \bm{X}$ and compared with the target to obtain the BER.

\begin{figure}
    \centering
    \includegraphics[width=.5\textwidth]{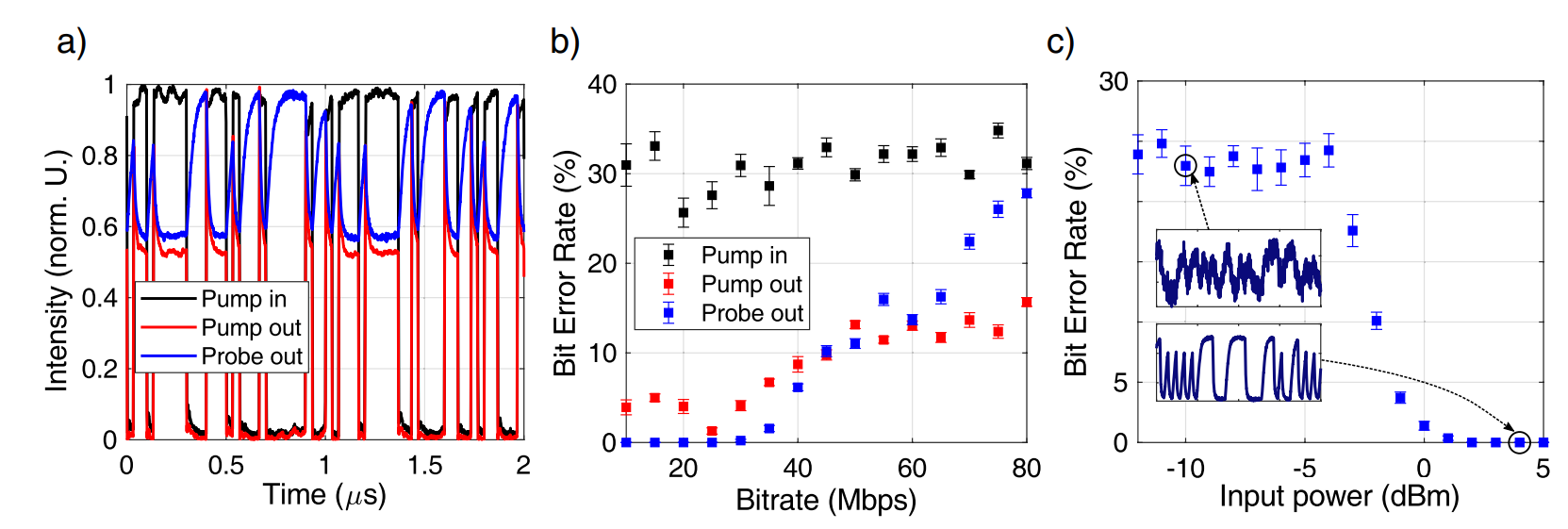}
    \caption{(a) Examples of waveforms processed during the XOR task. The pump laser driving the input port of the MR is shown in black, while the pump and probe outputs from the Drop port are respectively shown in red and blue. (b) Bit Error Rate (BER) as a function of the bitrate for the 1-bit delayed XOR task. Black dots use a predictor matrix $\bm{X}$ whose entries are sampled from the input pump power. Red and blue dots use respectively predictors sampled from the pump and the probe traces at the Drop port of the MR. In all three cases, the average pump power is set to 3 dBm. (c) Bit Error Rate as a function of the average pump power for a fixed bitrate of 20 Mbps. The insets show details of the probe waveform at the pump powers -10 dBm and 4 dBm. Figure from \cite{borghi2021reservoir}.}
    \label{fig:xorborghi}
\end{figure}

The results are shown in Fig. \ref{fig:xorborghi}. The traces reported in panel (a) illustrates the response of the MR to a binary input signal, highlighting the incoherent transfer of information between pump and probe signals driven by the nonlinear dynamics of the MR. The BER as a function of the input bitrate $B$ is reported in panel (b) and is evaluated in the case in which the state matrix $\bm{X}$ is sampled from the input pump signal (black dots) and the output pump (red dots) and probe (blue dots). In the first case, high BER values highlight the non-separability of the task, which cannot be solved at any bitrate. The nonlinear dynamics and the memory introduced by the MR determine an improvement in the performance of the RC network, especially at small bitrates. The BER increases again at high modulation frequencies since the free carrier dynamics become too slow compared to the pump power variations. Panel (c) reports the BER values for $B=20$ Mbps as a function of $P_{p}$, showing that the action of the reservoir becomes effective in BER reduction only when the nonlinear dynamics occur. \\

The second task on which the RC network has been trained consists in Iris flower recognition. The system receives in input the information (codified through a different $\bm{W}_{in}$) about the length and width of petals and sepals of a flower, with the objective of classifying it to one of the three possible species. Different configurations of the system are explored, reaching the highest recognition rate of $(99.3 \pm 0.2)\%$ for $N_v = 50$, $P_{p} = 7$ dBm, $B = 20$ Mbps and $\Delta = 50$ ns. In this configuration, $P_{p}$ is sufficiently high to trigger also thermal effects in the MR, which enter as a further element in the free carrier dynamics. The tests performed with the delayed XOR and Iris flowers classification have demonstrated the potentiality of a RC-NN approach realized with a MR resonator and linear classifiers. A key role in the RC network is played by memory and nonlinear dynamics in the MR, which increase the separability of the observables with respect to the virtual nodes. In the next section, further tests involving linear and nonlinear logic tasks performed on the RC network will be described.

\section{Linear and Nonlinear tasks on a Microring resonator}
\label{sec:V}

\begin{figure}[b]
	\centering
	\includegraphics[width=0.5\textwidth]{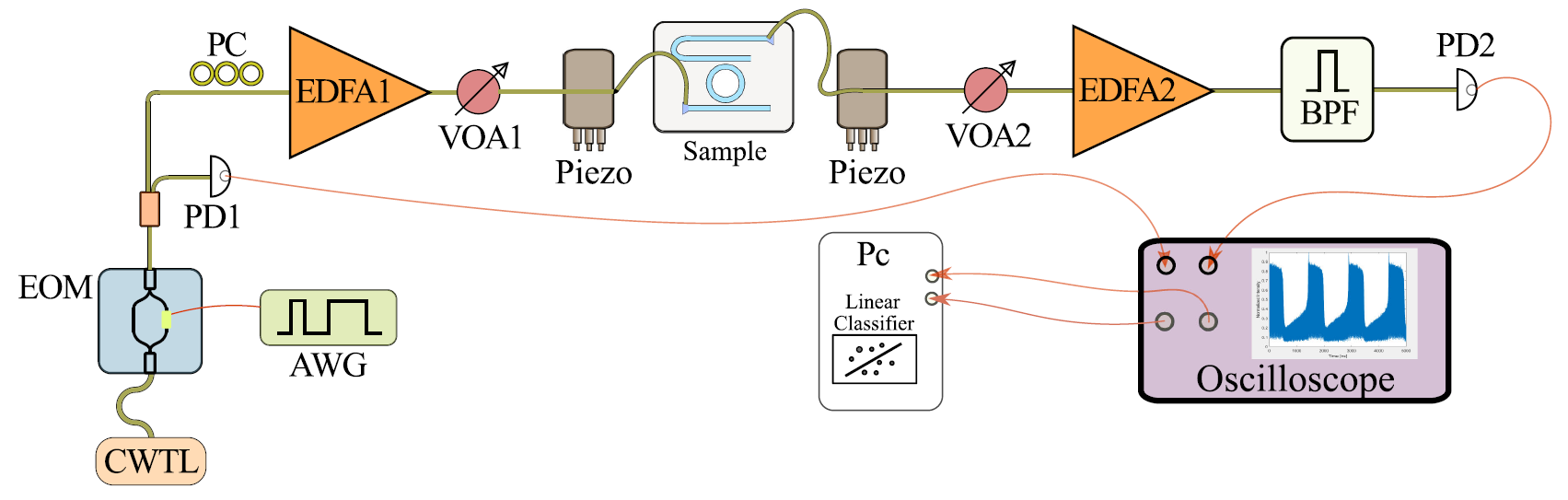}
	\caption{Diagram of the experimental setup. CWTL: Continuous Wave Tunable Laser, AWG: Arbitrary Waveform Generator, EOM: Electro-Optic Modulator, PD: photodetector, PC: polarization control, VOA: Variable Optical Attenuator, EDFA: Erbium Doped Optical Amplifier, BPF: Band Pass Filter, Pc: Personal computer. Note that the second amplification stage constituted by VOA2 and EDFA2 keeps the average power at the receiver at a constant value. This limits the variations of the Signal-to-Noise ratio (SNR) at PD2, since the most significant noise source at the detector is represented by thermal and shot noise. Adapted from \cite{bazzanella2022microring}.}
	\label{fig:bazz_setup}
\end{figure}

A MR resonator working in nonlinear regime serves the double purpose of providing memory to the system (on the typical time scale for FC relaxation and thermal cooling of the ring) and nonlinearities. To isolate these two mechanisms and to study their role in logic operations performed by a RC-NN, we investigated both linear logic tasks to get insights on the amount of fading memory induced by FC dynamics, as well as nonlinear logic tasks to access the interplay between memory and activation function \cite{bazzanella2022microring}. In this case, we used a simple MR to form a RC-NN together with the use of virtual nodes. The experimental setup, shown in  Fig. \ref{fig:bazz_setup}, is comparable with that of Fig. \ref{fig:borghi2setup}, except for the presence of the probe source only. Using the same formalism of the previous section, the input matrix to the system $\bm{X}_{in}$ is represented by a $1\times M$ binary sequence repeating in time, consisting in a PRBS sequence of order 8 and length $M=255$. The connectivity matrix $\bm{W}_{in}$ consists of a $N_v \times 1$ matrix of 1s, whose role is then simply to increase the dimensionality of the input matrix to $N_v \times M$. The input matrix is then $\bm{I} = \bm{W}_{in} \bm{X}_{in}$, where no rescaling or offset factors are applied so that each input virtual node $I_{i}^{j}$ has the same value (0 or 1) as the corresponding bit $x_{j}$ in the input sequence. The modulation imprinted in the pump laser corresponds exactly to $\bm{X}_{in}$, with $N_v$ virtual nodes in each bit and a total bit width of $T = N_v \delta $, with $\delta$ being the time duration of a single input virtual node. The virtual nodes at the output are obtained with the same modalities described above, and so it is for the training procedure. Here the output is collected directly from the pump, without relying on a probe signal.\\

\begin{figure}
    \centering
    \includegraphics[width =0.35\textwidth]{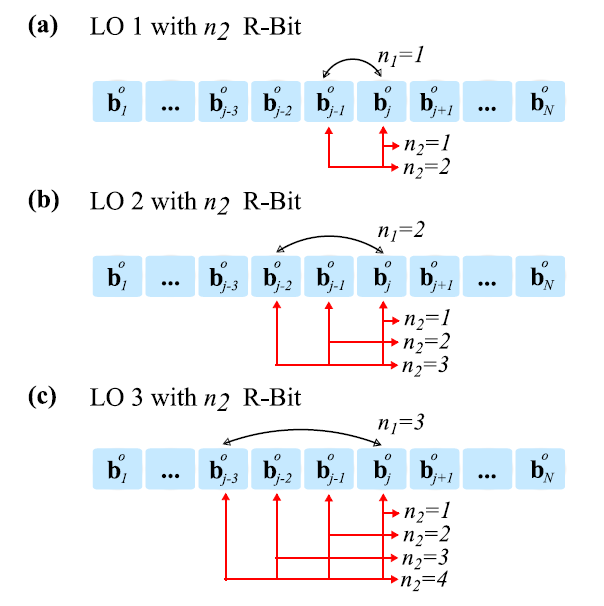}
    \caption{Sketch representing the three cases on which we tested the logical operations. According to the notation used in the text, $\bm{b}_j^0 = x_j$ and $N$ is the length in terms of bits of an input sequence, constituted by $N/M$ copies of $\bm{X}_{in}$ and $M=255$ is the length of a single $\bm{X}_{in}$ sequence (PRBS). $n_1$ indicates the distance between the bits on which the logical operation (LO) is performed and $n_2$ is the number of bits provided to the ridge regression in the training procedure. Note that the flow of bits is such that the past bits $\bm{b}^{i}_{j-n_1}$ are processed by the microresonator before the present bit $\bm{b}^{i}_{j}$. Figure from \cite{bazzanella2022microring}.}
    \label{fig:bazzanella_tasks}
\end{figure}

The tasks on which the RC-NN has been trained have been defined starting from both linear (AND, OR) and nonlinear (XOR) logic operations performed between the present bit and the $n_1$-th bit in the past of the input sequence $\bm{X}_{in}$. The target sequence consists then of a sequence of 0s and 1s with one single value for each input bit and it is obtained by applying the selected logic operation directly to the digitized input sequence. The linear classifier receives as input the state matrix $\bm{X}$ populated with virtual nodes associated with the current and previous bits up to $n_2$ bits (the ridge regression bits, or R-bits) in the past (for a total of $n_2 \times N_v$ virtual nodes). It produces a $1\times M$ output sequence $\bm{Y} = \bm{W_{out}} \bm{X}$ from which a digital sequence is obtained by the application of a threshold. A task is thus defined by the logic operation and the parameters $n_1$ and $n_2$, e.g. we refer to the AND operation with $n_1=1$ and $n_2=2$ as the AND 1 with 2 R-bits. The specific tasks on which the network is trained are obtained as variants of these basic operations, varying $n_1$ and $n_2$ as indicated in Fig. \ref{fig:bazzanella_tasks}. A mapping of the system in terms of state matrix is performed, varying the input bitrate, the average pump power entering the MR, and the detuning with respect to the resonant frequency of the MR. The training procedure is repeated for each of the so-obtained state matrices and also on those obtained sampling directly the input sequence, providing feedback on the effectiveness of the action of the reservoir. 

For the construction of each state matrix, the input and output curves are acquired with a fixed sampling rate of 20 GSa/s. The number of samples per bit $N_{s}$ varies then depending on the selected bitrate $B$, ranging from 20 Mbps to 4000 Mbps. How the virtual nodes are evaluated depends on the chosen number of virtual nodes $N_v$ in relation to the bitrate. Considering, for example, $N_v = 10$:
\begin{itemize}
	\item for $B<20$ GHz, one has $N_{s}>N_{v}$, thus the acquired samples within a single bit are grouped into $N_{v}$ bins and for each group, the average is performed;
	\item for $B=20$ GHz, one has $N_s = N_v$, thus the virtual nodes simply coincide with the acquired samples in each bit;
	\item for $B>20$ GHz, one has $N_{s}<N_{v}$, thus the first $N_s$ virtual nodes coincide with the acquired samples for that bit, while the remaining are set to zero.\\
\end{itemize}

For each analyzed task, the results are enclosed in a 3-graph figure, where the outcomes of the training procedure are accessed by means of the BER. In the first contour plot, each point represents the best value of the BER ($BER_{out}^{b}$) obtained with the RC-NN for a specific bitrate and detuning. Red points indicate the configurations in which the statistical limit is reached. The lowest input power that allows achieving $BER_{out}^{b}$ is presented in the second contour plot. Finally, the third plot shows the comparison between the results of the training procedure applied on the output of the RC-NN and directly on the input sequence (producing $BER_{in}^{b}$). The plot presents the ratio $RB = BER_{in}^{b}/BER_{out}^{b}$, with the red dots indicating where the statistical limit is reached in BER evaluation at the output (cross) or at the input (empty circles). Two color-maps are used: the gray scale describes the configurations where the RC-NN worsens the performance compared to the unprocessed optical sequence, while the contrary applies to the colored regions.  Black regions indicate an equal performance, namely $RB = 1$. \\

\begin{figure}
    \centering
    \includegraphics[width = 0.5\textwidth]{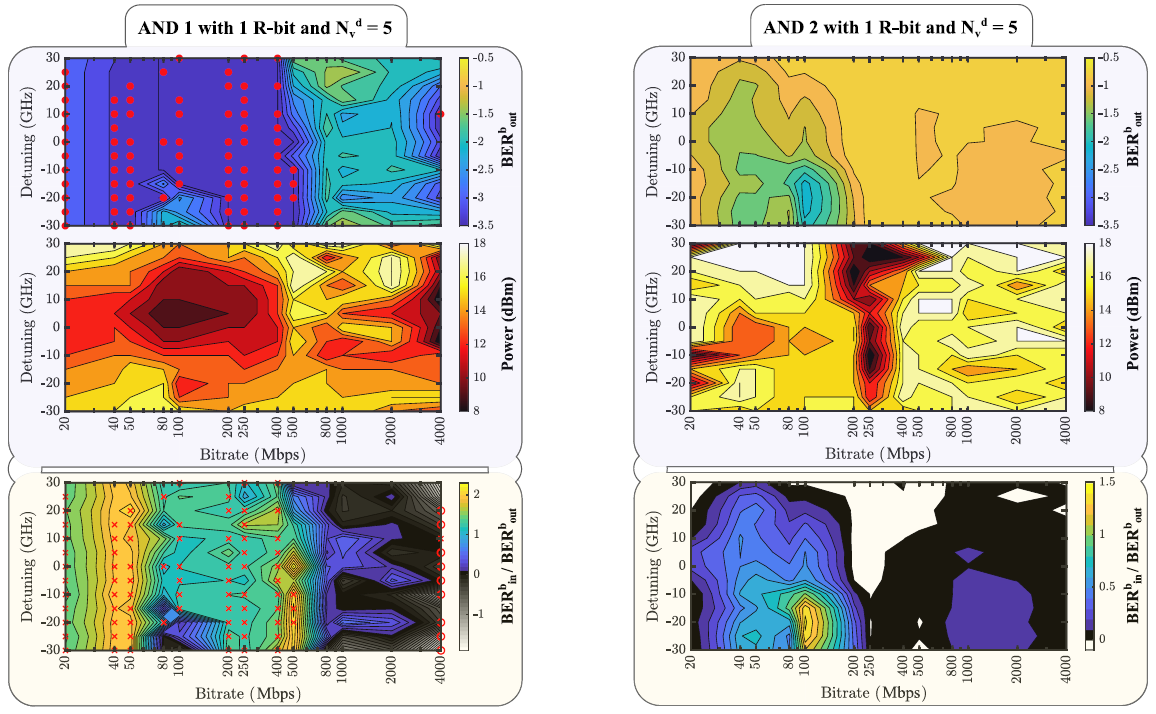}
    \caption{Maps as a function of the frequency detuning and input bitrate for AND 1 with 1 R-bit and $N_{d}^v=5$ (left column) and for AND 2 with 1 R-bit and $N_{d}^v=5$ (right column). (top panels) BER estimation from the RC-NN at the power which ensures the best network performances; (middle panel) the power at which the $BER_{out}^{b}$ values in the first panel are achieved; (bottom panel) the ratio between $BER_{in}^{b}$ and $BER_{out}^{b}$. All the values are given in a logarithmic scale. Figure from \cite{bazzanella2022microring}.}
    \label{fig:bazz_and}
\end{figure}

First, we focus on the solution of delayed linear tasks: the MR should provide memory to the system, temporarily storing the information of the input sequence in the nonlinear dynamics of the FC. The results for the AND 1 with 1 R-bits are presented in Fig. \ref{fig:bazz_and} (left). The statistical limit in $BER_{out}^{b}$ is reached in a vast region of the detuning-bitrate space, up to a bitrate of 500 Mbps. The second contour plot highlights the presence of a region in correspondence of a zero detuning where the task is resolved even for low input power. The third plot shows that lower BER values are reached with the optical processing operated by the MR compared to those obtained by applying the training to the input sequence, with the most evident benefits appearing in the region between 40 Mbps and 50 Mbps. This is the region where the MR nonlinearity provides enough memory to the RC-NN to solve the delayed logical operation. On the other hand, for high bitrates the performances associated with the two treatments tend to be comparable, observing also worsening of the performance induced by the MR for 4000 Mpbs. Indeed, for such a bitrate the statistical limit is reached for multiple detunings applying the training directly to the unprocessed input sequence. In this extreme case, the source of the memory is represented by the non-idealities introduced by the generation and detection stages, due to their limited electronic bandwidth. The nonlinear dynamics of the MR do not provide further memory, on the contrary, it causes distortions in the signal, assuming thus a detrimental role in the training process. This is also highlighted by the darker region present on the power map for high bitrates, asserting that the corresponding $BER_{out}^{b}$ values are obtained with minimum input power, thus trying to minimize the nonlinear effects induced by the MR.

Figure \ref{fig:bazz_and} (right) presents the results for AND 2 with 1 R-bit, for which a higher amount of memory is required. The BER map in the top panel shows a region in correspondence of a detuning of 20 GHz and a bitrate of 100 Mbps where the $BER_{out}^{b}$ are lower compared to the rest. The power map shows that the best results are obtained for high input power values, which ensures the presence of nonlinear effects in the MR. In this configuration of bitrate and detuning, the action of the MR improves the performance of the linear classifier, compared to the training performed on the optically unprocessed input sequence. This aspect is highlighted in the bottom panel, where the higher RB ratio appears in the same map region and amounts to $RB = 10^{1.5}$. The fact that the lowest $BER_{out}^{b}$ value appears in correspondence with a negative detuning suggests that the nonlinear dynamics occurring in the MR are related to FC rather than thermal relaxation \cite{Borghi2021modeling, johnson2006selfinduced}. The training procedure performed on the AND 3 with 1 R-bit returned a minimum $BER_{out}^{b}$ of $10^{-1}$, demonstrating that the memory of the MR related to its internal nonlinear dynamics is limited to 2 bits. \\
 
\begin{figure}
    \centering
    \includegraphics[scale=.5]{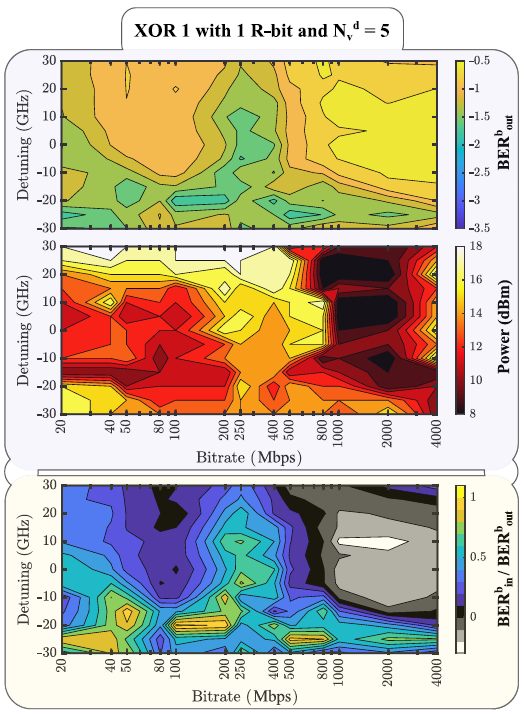}
    \caption{Maps as a function of the frequency detuning and input bitrate for XOR 1 with 1 R-bit and $N_{d}^v=5$. (top) BER estimation from the RC network at the power which ensures the best network performances; (middle) the power at which the $BER_{out}^{b}$ values in the first panel are achieved; (bottom) the ratio between $BER_{in}^{b}$ and $BER_{out}^{b}$. All the values are given in a logarithmic scale. Figure from \cite{bazzanella2022microring}.}
    \label{fig:bazz_xor}
\end{figure}

The network has been also trained to solve the XOR 1 with 1 R-bit. In Fig. \ref{fig:bazz_xor}, the top panel contour plot shows various regions of low $BER_{out}^{b}$, reaching a minimum value of $10^{-1.7}$. Many local minima lie in the negative detuning half-plane, where the combination with high input power triggers nonlinear effects related to FC dynamics in the MR. On the contrary, in correspondence with positive detuning and high input power, the system sees a performance degradation, a symptom that nonlinearities induced by thermal effects are detrimental. It is interesting to notice that the task is solved in the region around the bitrate of 250 Mbps, which is the inverse of the typical FC lifetime in the MR ($\tau_{fc} \sim 4.5$ ns and $\tau_{th} \sim 100$ ns). The higher BER values in the map appear for bitrates higher than 800 Mpbs. In the same region of the power map, low input power values are registered, meaning that the system tries to avoid entering the nonlinear regime in the MR. Even if the RC-NN does not manage to perfectly solve the task in any of the explored conditions, it still provides performance enhancement compared to the training performed on the unprocessed optical input. Indeed, the colored regions in the RB contour plot cover the majority of the map, corresponding to the regions where low BER values are observed in the BER map. With the RC-NN, nonlinearities generate memory in the system and act also as an activation function, providing effective separability of the virtual nodes for the linear classifier. On the contrary, for high bitrates the intersymbol interference induced by the modulation process becomes more significant, thus inserting a sufficient amount of memory already into the input signal. Subsequent nonlinear processes in the MR become unnecessary.\\

The results illustrated in this section witness the role of the MR in providing both memory and nonlinearity. The tests performed on the structure used as a reservoir for linear delayed tasks revealed a maximum amount of memory of 2 bits provided by the MR. On the other side, both memory and nonlinearity induced in the MR are necessary to solve nonlinear delayed tasks, revealing to be detrimental when the combination of fast modulation and detection already induces sufficient separation in the virtual nodes. As we saw in the current and the previous sections, MRs are versatile tools whose properties in terms of memory and nonlinearities can be used to solve logic and analog tasks. A single MR has, however, some limitations. One of the most evident is related to the restricted amount of memory that it provides. 

\section{Microring resonators with external optical feedback for time delay reservoir computing}
\label{sec:VI}

Away from the self-pulsing regime, the amount of memory provided by a MR is linked to the typical lifetime of FC and the thermal relaxation of the structure, both of which are larger than the typical photon lifetime in the MR. An external optical feedback inserted in the structure enhances the MR memory and allows the use of MR-based RC-NN in time series prediction \cite{Donati22microring}. 
\begin{figure}
	\centering
	\includegraphics[width=.5\textwidth]{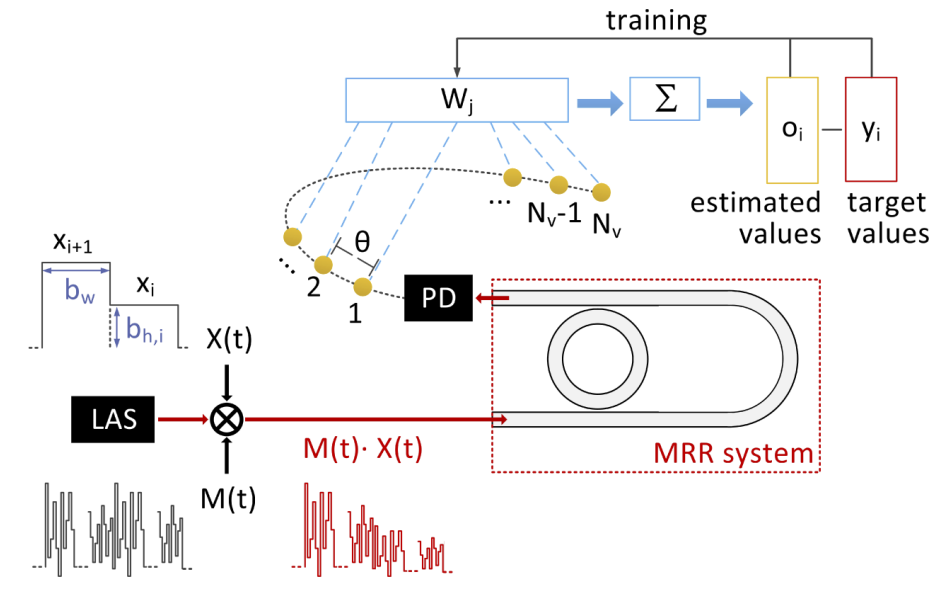}
	\caption{Schematic of time delay RC-NN with a MR subject to optical feedback. The MR structure is shown in the dashed box. It is in an add-drop configuration with the external optical feedback where both the phase $\phi$, the amplitude $\eta$ and the dealy $\tau_F$ can be controlled. $\gamma_e$ represents the MR extrinsic losses due to the coupling with the straight waveguides. The encoded information $X(t)$ is masked with a sequence $M(t)$ and modulates the optical power from the laser (LAS) emission. At the drop port, the photodetected (PD) signal provides the time-multiplexed output states of the reservoir, which are weighted and linearly combined to compute the predicted value $o_i$. The weight optimization is performed via a linear classifier, with supervised learning over the expected values $y_i$ data set. Figure adapted from \cite{Donati22microring}.}
	\label{fig:donati}
\end{figure}

The scheme of a simulated RC-NN based on a MR with external optical feedback is presented in Fig. \ref{fig:donati}. The MR operates in an add-drop configuration with two equal coupling coefficients $\gamma_e$ to the bus waveguides, as sketched in the dashed box. The signal is coupled to the MR from the input port (left-bottom) while the output signal is collected from the Drop port (top right). The Through (bottom right) and Add (top right) ports are coupled by a feedback loop providing a delay of $\tau_F$ and a tunable phase shift $\phi_F$. The feedback strength is controlled by $\eta_F \in [0,1]$, with $\eta_F=0$ standing for full attenuation ($E_{add}=0$) and $\eta_F = 1$ standing for null attenuation. Practically, the feedback loop can be realized by an optical fiber, a variable optical attenuator, a SOA (to recover the insertion losses), and a phase shifter. Using a scattering matrix approach, the electric field amplitude at the Through ($E_{th}$), Drop ($E_{drop}$), and Add ($E_{add}$) ports are related to the input field ($E_{inp}$) by:
\begin{align*}
    &E_{th}(t) = t_r E_{inp}(t) + \sqrt{2 \gamma_e} U(t) \\
    &E_{drop}(t) = \sqrt{2\gamma_e} U(t) + t_r E_{add}(t) \\
    &E_{add}(t) = \sqrt{\eta_F} \text{e}^{-i\phi_F} E_{th}(t-\tau_F)
\end{align*}
where $t_r$ indicates the transmission coefficient respectively from the Input to the Through port and from the Add to the Drop port, while $U(t)$ represents the optical energy amplitude circulating in the MR. The time dependence of $U(t)$ derives from the interplay between linear and nonlinear dynamics in the MR, as described in section \ref{sec:III} and in \cite{Borghi2021modeling}. Typical values for MR parameters are a Q-factor of $Q=3.19\times10^{4}$, an intrinsic photon lifetime $\tau_{ph} \sim 50$ ps, a free carrier lifetime $\tau_{FC} \sim 3$ ns and a thermal lifetime $\tau_{TH} \sim 83$ ns.

To create the input optical signal (Fig. \ref{fig:donati} bottom left), we start from an analog or digital sequence $X(t)$, with every bit $x_i$ having a duration of $b_w$ (common to all the bits) and an amplitude of $b_{h,i}$. A periodic random mask $M(t)$ is then applied to the sequence in order to increase the dimensionality of each bit to $N_v$ (namely the number of virtual nodes chosen for the reservoir), corresponding to a time-separation between mask values of $\theta=b_w/N_v$. Each mask entry is sampled from a uniform distribution and the periodicity condition $M(t) = M(t + b_w)$ is obeyed. The so-obtained masked sequence $X(t)M(t)$ is used as a modulation pattern for a CW optical signal provided by a laser operating at a given optical power $P_{max}$ and detuning $\Delta \lambda_s$ with respect to the linear resonance frequency of the MR. The resulting input optical field is then written as 
\begin{equation}
    E_{inp} (t)= \left[ X(t) M(t) \right]^{1/2} = \left[ x_i m_j \right]^{1/2} , 
\end{equation}
for $b_w(i-1) + \theta(j-1) \leq t \leq b_w(i-1) + \theta j$. The sequence of virtual nodes $N_{j,i}$ with $j=1,\dots,N_v$ associated with the input bit $x_i$ is obtained from the detected output signal $\vert E_{drop} \vert^2$. The output virtual node $N_{j,i}$ represents the sample of the output signal acquired simultaneously with the injection of $x_i m_j$ in the structure, in formulas
\begin{equation}
    N_{j,i} \propto \vert E_{drop} (b_w (i-1) + \theta j) \vert^2 .
\end{equation}
A unique estimator $o_i$ for each sequence $N_{j,i}$ with $j=1,\dots,N_v$ (namely for each input bit $x_i$) is produced as 
\begin{equation}
    o_i = \sum_{j=1}^{N_v} W_j N_{j,i}.
\end{equation}
where $W_j$ is a $N_v$ dimensional vector. The training of the network is performed by linear regression, which provides the values $W_j$ that minimize the Normalized Mean Square Error (NMSE) between the predictions $o_i$ and the nominal outcomes $y_i$ obtained from $x_i$ by means of a given task. For each selected operation, the performance of the network has been accessed through a mapping of the NMSE values as a function of $P_{max} \in $ [1,8] mW, $\Delta \lambda_s \in $ [-50,50] pm, $\eta_F \in $ [0,1] and $\phi_F \in $ [0,2$\pi$]. The numerical experiments have been performed with $\theta = 40$ ps, meaning that, since $\theta \approx \tau_{ph}$, photons circulating in the ring contribute to short-term memory creation. The bit width is set to $b_w = 1 \;\text{ns} \approx \tau_{FC}$ ($N_v = 25$), so that each input bit $x_i$ manages to produce observable variations in the FC population and triggers nonlinear effects which cause in turn longer-term memory formation. Finally, the delay introduced by the feedback loop amounts to 1 ns.\\

Along with the analysis of the performance of the reservoir computing approach in solving specific tasks, a study on the amount of memory achieved by the system in specific conditions has been conducted. The reservoir is provided with a random sequence of bits sampled from a uniform distribution and it is trained to remember the $l$-th previous element of the input sequence. The amount of memory for the trained system is obtained by evaluating the Memory Capacity (MC)
\begin{equation}
    MC = \sum_{l=1}^{l_{max}} m(l), 
\end{equation}
with
\begin{equation}
m(l) = \frac{cov^2(o(n), i(n-l))}{\sigma_o^2 \sigma_i^2},
\end{equation}
where each term $m(l)$ measures the covariance ($cov$) between the output vector $o(n)$ and the input bit sequence $i(n-l)$ delayed by $l$ bits, with $\sigma_o$ and $\sigma_i$ representing the respective variances. The influence of nonlinear processes in memory formation can be detected by simulating the evolution of the standard deviation of the wavelength resonant shift $\sigma (\lambda_0)$: high values are symptom of the presence of nonlinear effects. These numerical simulations for the dynamics of the MR have been performed by integrating the 3 canonical coupled differential equations derived from the couple mode theory \cite{Donati22microring,johnson2006selfinduced}. \\

\begin{figure}
    \centering
    \includegraphics[width=.5\textwidth]{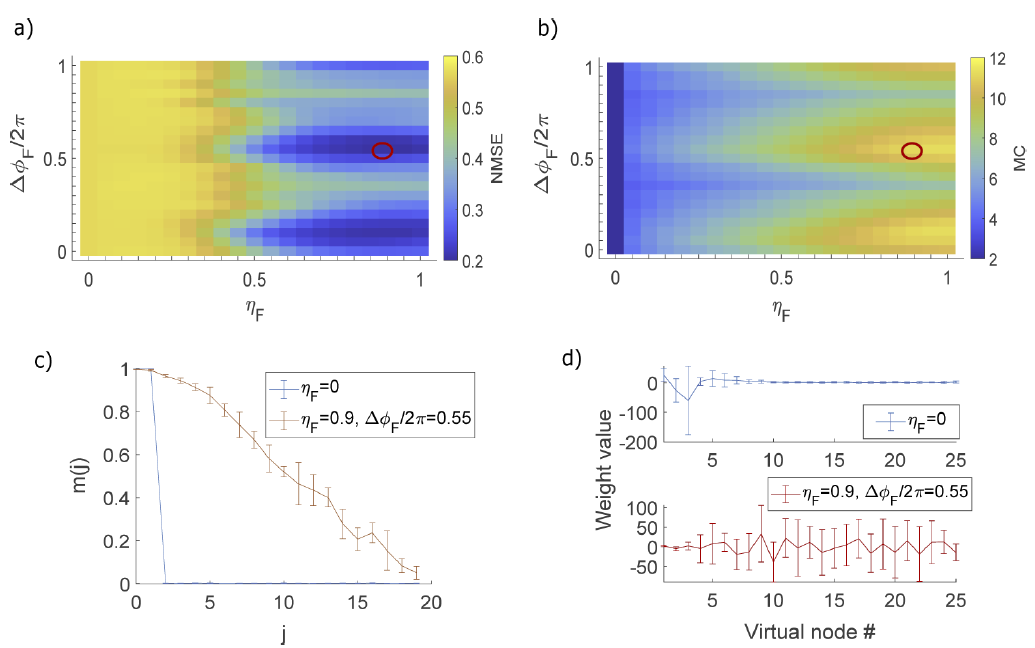}
    \caption{Performance of the RC-NN for the Narma-10 benchmark task. (a) NMSE and (b) MC, versus optical feedback strength $\eta_F$ and phase $\Delta \phi_F$. Red circle denotes the conditions with the lowest NMSE. (c) Memory function $m(l)$, for the cases without feedback (blue line) and with feedback conditions that result in the lowest NMSE (red line). (d) Readout weights for a task to remember the previous input value $x_{i-1}$, for the cases without feedback (blue line) and with feedback conditions that result in the lowest NMSE (red line). MC is computed using $l_{max} = 19$. The initial wavelength shift is $\Delta \lambda_s = -10$ pm and the MR is operating in the linear regime, with $b_w = 1$ ns. Figure from \cite{Donati22microring}.}
    \label{fig:donatiNarma10}
\end{figure}

The first benchmark task is represented by the Narma-10 task, in which the system is trained to predict the response of a discrete-time tenth-order nonlinear auto-regressive moving average \cite{jaeger2002adaptive}. In order to solve the task, it is necessary for the system to show a memory of at least 10 bits. The results are reported in Fig. \ref{fig:donatiNarma10}. Panels (a) and (b) contain respectively the NMSE and the MC parameter values mapped as a function of the feedback parameters $\eta_F$ and $\phi_F$, for the optimized values $P_{max} = 0.1$ mW and $\Delta\lambda_S = -10$ pm. Red circles highlight the configuration for which the NMSE is at its minimum value, which is obtained in a region where the MC parameter approaches its maximum measured value. The best performance of the system is thus obtained when operating in a linear regime and considering a strong feedback signal ($\eta_F = 0.9$). In this configuration, the memory necessary to solve the task is provided by the external feedback loop, without the need for the MR nonlinearities as a further source of memory. The memory increase provided by the feedback loop is evident in Panel (c): the system without the introduction of the external feedback shows a maximum MC parameter of 2, compared to a maximum value of about 13 in the optimal feedback configuration. Indeed, with $\eta_F = 0$ and low input power, the only source of memory is related to photon lifetime in the ring: information retained from the past bit is only related to the inertia between the last virtual nodes of the previous bit $x_{i-1}$ and the first ones of the current bit $x_i$. This result is also deducible from Panel (d), which presents the values of the weights found by the RC classifier for the system in the memory-less and optimal configuration, respectively. For $\eta_F = 0$, only the weights related to the first virtual nodes are significant compared to the others, while in the optimal feedback configuration, all the virtual nodes assume a role in carrying information.\\

\begin{figure}
    \centering
    \includegraphics[width=0.5\textwidth]{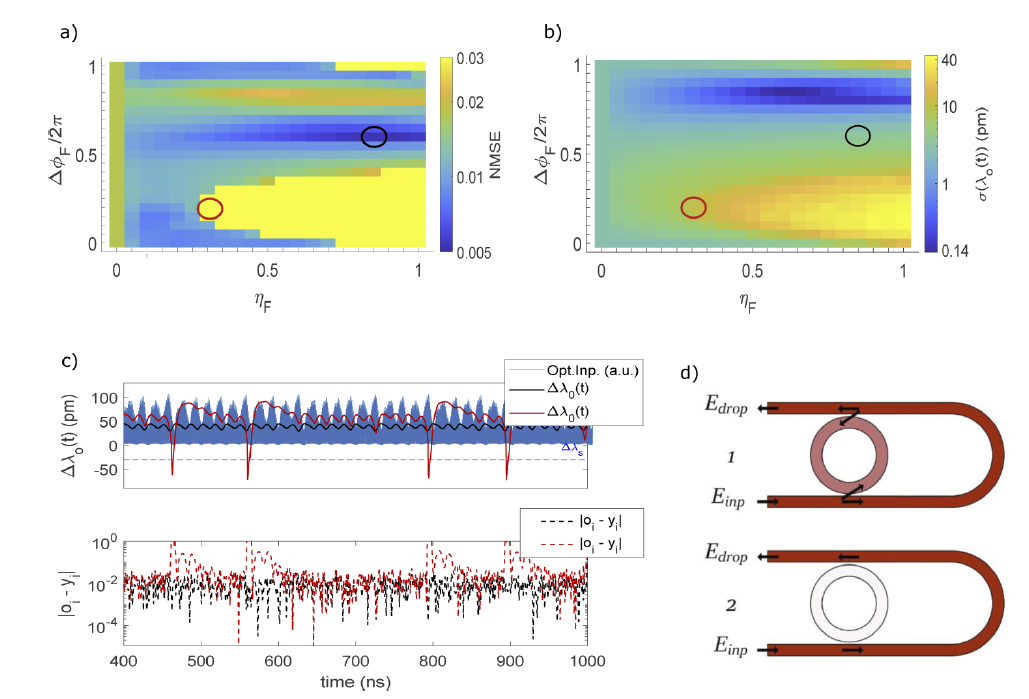}
    \caption{Performance of the RC-NN for the Mackey-Glass benchmark task. (a) NMSE and (b) standard deviation of the resonance wavelength shift $\sigma(\lambda_0)$, versus optical feedback strength $\eta_F$ and phase $\Delta \phi_F$ of the MR system. Black (red) circle denotes the conditions with the lowest (highest) NMSE. (c) Temporal evolution of the resonance shift and the bit error $\vert o_i - y_i \vert$ during the task for two feedback conditions: the black line corresponds to the lowest NMSE (black circle, (a)), and the red line corresponds to the highest NMSE (red circle, (a)). (d) Dynamical operation of the MR with optical feedback under self pulsations: light occasionally enters (path 1, upper) or bypasses (path 2, lower) the MR. The initial wavelength shift is $\Delta \lambda_S = - 30 $ pm, the maximum launched optical power at the input is $P_{max} = 5$ mW and $b_w = 1 $ ns. Figure from \cite{Donati22microring}.}
    \label{fig:donatiglass}
\end{figure}

Another benchmark test is represented by the Mackey-Glass prediction task which we operated in a weakly chaotic behavior \cite{jaeger2004harnessing}. In this case, the RC-NN is required to predict the next bit $x_{i+1}$ knowing the current bit $x_i$. An overview of the performances reached by the system is portrayed in Fig. \ref{fig:donatiglass}. Panel (a) and (b) report the NMSE and $\sigma (\lambda_0)$ as a function of $\eta_F$ and $\Delta \phi_F$, while keeping $P_{max} = 5$ mW and $\Delta \lambda_S = -30 $ pm (optimal operational conditions). The black circle highlights the configuration corresponding to the lowest NMSE overall: the high value of $\sigma(\lambda_0)$ of Panel (b) indicates that, contrary to what is shown for the Narma-10 task, here the optimal configuration is obtained by exploiting nonlinearities induced in the MR. Notice also that the feedback strength is large, but the signal reaching the Add port is in an intermediate state between constructive and destructive interference with the signal circulating in the MR. Indeed, a fully constructive interference condition would have promoted even more significant variations in $\lambda_0$, but more nonlinearities would have been detrimental to the performance (Panel (a)). In this case, the feedback loop has the double purpose of extending the memory and tuning the level of nonlinearities induced in the MR. The worst performance of the system is highlighted in Panel (a) and (b) by the red circles. In Panel (c) this configuration is shown by the red curves. The deep spikes in $\Delta \lambda_0 (t)$ appearing in Panel (c) are symptoms of the presence of SP in the MR, which degrades the performance. When operating in SP, the light does not couple with the MR, but propagates straight through the delay line and then to the Drop port, without recirculating in the MR (Panel (d), configuration 2).

In conclusion, the delay feedback loop coupled with the MR has proven effective as a memory extender for the RC-NN. When operating in a linear regime, the MR coupled with the delay loop assumes the function of a shift register in the optical domain, while when nonlinearities are triggered in the MR, the delay loop serves the purpose of both controlling the strength of nonlinearities and extend the memory of the system. 

\section{Conclusions}
\label{sec:VII}

Different optical circuits have been described as basic elements of PNNs within a silicon photonics platform. Multiple architectures have been analyzed and tested with respect to various tasks, demonstrating their specific properties and establishing performance benchmarks. A simple delayed complex perceptron employed as a feed-forward neural network with memory has proved effective in compensating distortions induced by chromatic dispersion in a 10 Gbps NRZ signal propagating in a 125 km fiber. The trained perceptron restores the opening of the eye diagram of the signal after the propagation, thus drastically diminishing the BER compared to the uncompensated signal at the fiber output. The optical processing operated by the perceptron permits minimizing latency and tuning the properties. In addition, adding more delay lines and nonlinearities would increase the computational capabilities of the complex perception, leading thus to more advanced mitigation actions in optical communications \cite{argyris2022photonic}.

Furthermore, the use of passive elements in PNNs is of extreme importance to reduce the power budget. Interesting possibilities are given by the nonlinear dynamics of a MR. These have been explored, observing the self-pulsing regime in different configurations of input power and initial detuning. Nonlinear dynamics plays a fundamental role in long short-term memory formation in the MR and enables the use of a MR as a reservoir in a RC-NN. In one implementation, the FC dynamics triggers the incoherent transfer of information between pump and probe signals, increasing the separability of the data which allows the use of linear classifiers to achieve complex tasks. Remarkably the best performance is achieved at the edge of the self-pulsing regime where both the free carrier dispersion and the thermo-optic effect are critical.
Out of the pump-and-probe approach, linear delayed tasks have been adopted to decouple the memory formation from the role of the MR used as a nonlinear node. These tests revealed the finite memory retained by the MR, which is limited to 2 bits at the best bit rate. The introduction of an external feedback loop coupled with the MR represents an effective memory source. In the RC-NN approach with virtual nodes, this new structure is able to time series forecast with memory on timescales larger than those typically associated with nonlinear processes induced by FC dynamics (not thermal effects) in a MR. 

Here we have discussed a few examples of the use of single MR in PNNs. More elaborated and complex neural networks are possible when matrices of perceptrons or microrings are used, as reviewed in \cite{markovic2020physics, sunny2021survey,brunner2021competitive, shastri2021photonics}. PNNs with extremely high performances and speed have been demonstrated in \cite{xu202111,feldmann2021parallel}.

\begin{acknowledgments}
This project has received funding from the European Research Council (ERC) under the European Union’s Horizon 2020 research and innovation programme (grant agreement No 788793, BACKUP and No 963463, ALPI) and from the MUR under the project PRIN PELM (grant number 20177 PSCKT).
\end{acknowledgments}

\bibliography{bibliography}

\end{document}